\documentclass[12pt]{article}
\usepackage{graphicx}
\usepackage[cp1251]{inputenc}
\usepackage{rotating}
\begin{document}

 \vskip 0.5cm
  \centerline{\bf\large Quiet Region Jet by Eruption of Minifilament }
  \centerline{\bf\large  and Associated Change in Magnetic Flux}
 \bigskip
 \bigskip
  \centerline
 {
 Rakesh Mazumder
 }
 \bigskip
 \centerline{\small \it
  Center of Excellence in Space Sciences India, Indian Institute of Science Education and Research Kolkata,}
 \centerline{\small \it Mohanpur 741246, West Bengal, India, E-mail: rm13rs027@iiserkol.ac.in}
 \bigskip
 \bigskip
 \bigskip


{
{\bf Abstract}--- We observe a coronal jet around 22:08 UT on 23$^{rd}$ March of 2017, in a quiet region towards the north-east of the solar disk. A minifilament eruption leads to this jet. We analyze dynamics of the minifilament from its formation until its eruption. We observe that the minifilament starts forming around 15:50 UT. Then at a later time, it disappears but again appears and finally erupts as a blowout jet around 22:08 UT. We study the evolution of photospheric magnetic field beneath the minifilament. Initially, we observe subsequent increase and decrease of positive magnetic flux. Prior to the initiation of the eruption, positive magnetic flux shows a continuous decrease until the minifilament disappears. The positive magnetic flux increases and decreases due to new positive flux emergence and cancellation of positive magnetic flux with negative magnetic flux respectively.
}



\section{Introduction}
Coronal jets are collimated transient features in solar corona [23]. They are usually observed in extreme ultraviolet (EUV) [35] and X-ray [2,3,29] emission. Coronal jets show a brightening at the edge of their base and are known as jet based bright points (JBPs). Jets are usually observed at the region of magnetic flux emergence or magnetic flux cancellation  [22,28,31].
The occurrence of coronal jet due to eruption of minifilaments is reported by many authors [10,21,22,28,30,31].\\
Eruptive minifilaments are ubiquitous on solar disk [4,19,24,25,34,37]. The minifilaments have projected length of about 20 Mm [22,30,34]. The association of minifilament with magnetic flux cancellation is also reported [1,10,11,36].\\
 Similar to filaments, minifilaments are also always found in the vicinity of neutral line (also known as Polarity Inversion Line, PIL) [9,17]. Studies have found that minifilaments are miniature counter-part of large-scale filaments [9,30,34]. Hence understanding the formation mechanism of minifilaments can shed light on formation mechanism of large-scale filaments. Often big filaments erupt and give rise to coronal mass ejection (CME). Since CME is hazardous to space weather, study of filaments is important for space weather research [7,8,12]. \\
We organize this paper as follows: In section 2 we describe our data sources and data quality. In section 3 we describe data analysis and the results are also discussed. In the final section, we summarize and conclude our findings.

\section{Data and Observation}
A coronal jet is observed, and the evolution of its source (a minifilament) is studied using data from Solar Dynamical Observatory(SDO)/Atmospheric Imaging Assembly(AIA) in different wavelengths (171~\AA, 193~\AA \hspace{0.5mm} and 304 ~\AA) [14]. AIA gives the full image of the Sun in extreme ultraviolet(EUV) and Ultraviolet (UV) wavelength with the spatial resolution of the pixel size being $0.66 ''$ and temporal cadence of 12 seconds.  SDO/Helioseismic Magnetic Imager (HMI) [27] line of sight (LOS) magnetograms is used to study magnetic field evolution underneath the minifilament. HMI gives full sun LOS magnetograms with the resolution of the pixel size of $0.5 ''$ and 45 seconds temporal cadence [26].

\begin{figure}[htbp!]
\center{\includegraphics[scale=0.5,angle=90]{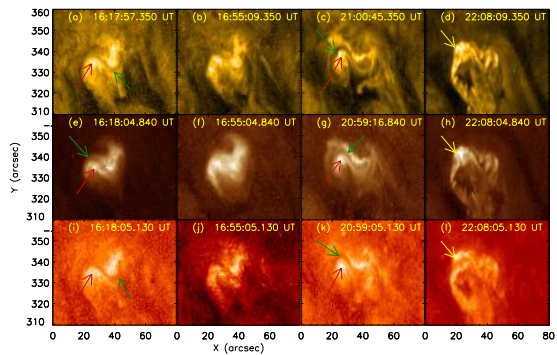}}
\caption{The temporal evolution of the minifilament is depicted in panel (a) to (l) in different wavelength channels. Panel (a) to (d) shows the temporal evolution of minifilament in AIA 171 ~\AA \hspace{0.5mm}. Panel (e) to (h) depicts the temporal evolution of minifilament in AIA 193 ~\AA \hspace{0.5mm}. Panel (i) to (l) shows the temporal evolution of minifilament in AIA 304~\AA \hspace{0.5mm}.}
\label{context}
\end{figure}

\begin{figure}[htbp!]
\center{\includegraphics[scale=0.6,angle=90]{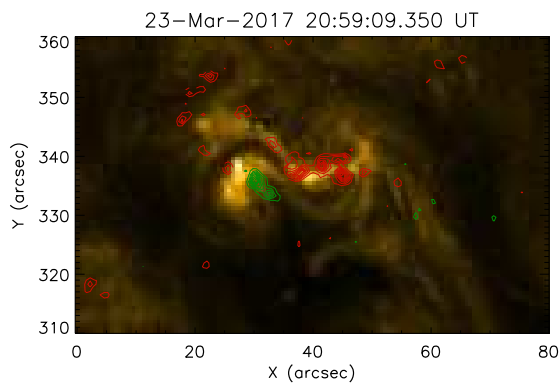}}
\caption{The image of minifilament in AIA 171 ~\AA \hspace{0.5mm}. The positive and negative contours of HMI magnetic field are overplotted in green and red respectively.}
\label{contour}
\end{figure}
\begin{figure}[htbp!]
\center{\includegraphics[scale=0.8,angle=90]{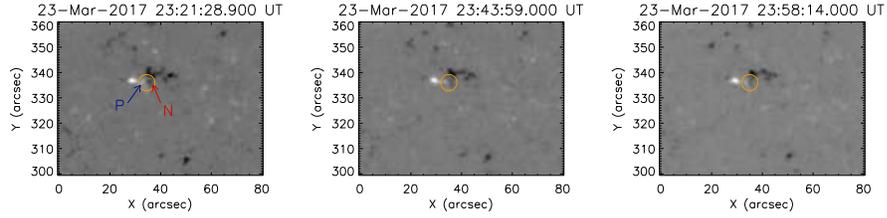}}
\caption{Evolution of the magnetic field in three images of HMI magnetogram are shown subsequently one after the other. We have encircled the region of interest with an orange circle. In left panel the positive patch of magnetic field P is marked with blue arrow whereas the negative patch of magnetic field N is marked with red arrow. However, in the middle and right panels, those encircled negative and positive magnetic patches almost disappear.  }
\label{example}
\end{figure}

\begin{figure}[htbp!]
\center{\includegraphics[scale=0.8,angle=90]{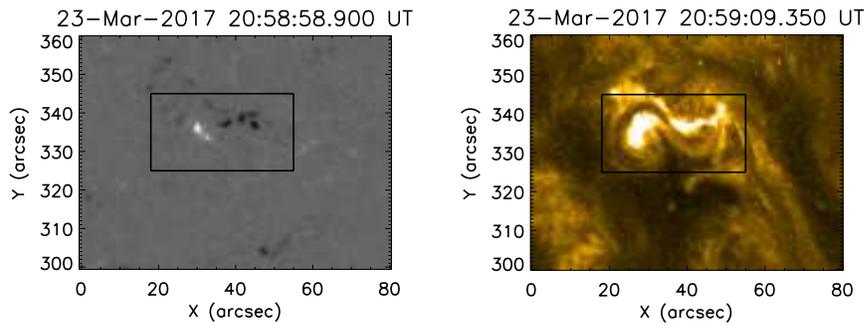}}
\caption{The left panel shows LOS magnetogram underneath minifilament. The black rectangular box shows our selection of the box for magnetic flux calculation. The right panel shows minifilament in AIA 171 ~\AA \hspace{0.5mm}. Our selected box is overplotted on it. }
\label{box}
\end{figure}

\begin{figure}[htbp!]
\center{\includegraphics[scale=0.5,angle=90]{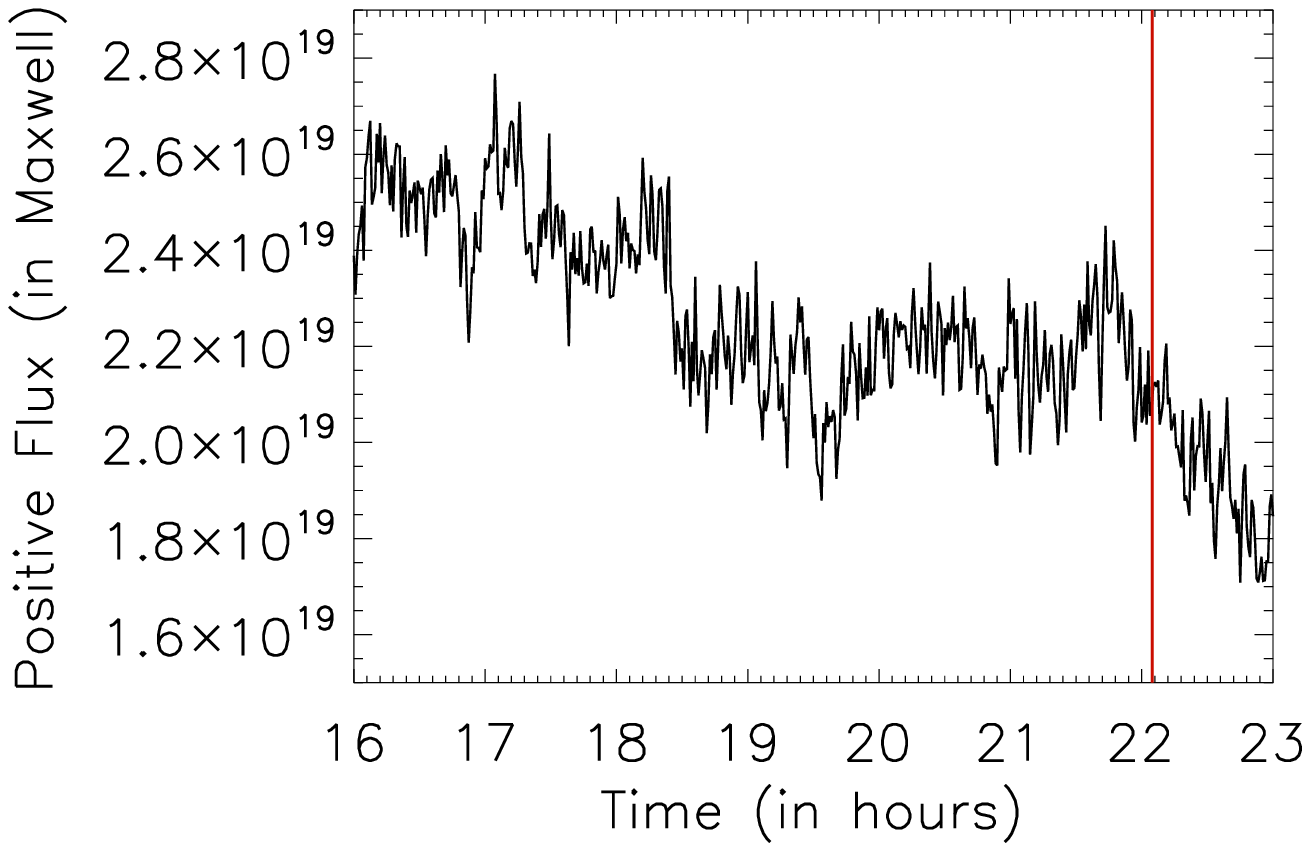}}
\caption{The evolution of positive magnetic flux inside our selected box (see Figure 4). The time starts from 2017 $3^{rd}$ March 16:00 UT. The red vertical straight line shows the time of initiation of eruption of the minifilament (approximately 22:08 UT). }
\label{flux}
\end{figure}

\section{Data analysis and Results}
\subsection{Overview}
We detect a blowout jet in a quiet region at north-east of solar disk around 22:08 UT on 23$^{rd}$ March of 2017. A minfilament eruption is found to be the source of the jet. We study the evolution of the minifilament in detail. AIA 171~\AA, 193~\AA \hspace{0.5mm}, and 304~\AA \hspace{0.5mm} wavelength channels are used to study the evolution of the minifilament. We also study the photospheric magnetic field evolution in the minifilament region using HMI LOS magnetograms.  
 All the images (both in AIA and HMI) are derotated using one of them as a reference image.

\subsection{Evolution of the Minifilament}
We carry out a detailed analysis of the minifilament in three extreme ultraviolet wavelengths (AIA 171, AIA 193 and AIA 304) using AIA onboard SDO.
The minifilament starts forming around 15:50 UT on 23$^{rd}$ March of 2017. We study its evolution from 15:50 UT to 23:00 UT. Figure 1 depicts the evolution of minifilament from its formation till initiation of eruption. Figure 1 [images (a) to (d)] depicts the evolution of the minifilament in AIA 171~\AA. We see in Figure 1.(a)
at time 16:17:57.35 UT the minifilament is formed and its spine is clearly visible which almost disappears at a later time 16:55:09.35 UT  (see Figure 1.(b)). Then at time 21:00:45.35 UT the minifilament spine reappears (see Figure 1.(c)). Finally around 22:08 the minifilament is seen to initiate its eruption (see the Figure 1.(d)). 
 Figure 1 [images (e) to (h)] and Figure 1[images (i) to (l)] shows same type of the evolution of minifilament in AIA 193~\AA \hspace{1mm} and AIA 304~\AA \hspace{1mm} respectively. In the first and third column of Figure 1 we marked the minifilament spine and nearby brightening by green and red arrows respectively. We observe brightening near minifilament spine (marked by red arrows in first and third column of Figure 1) when minifilament is visible. However, when minifilament disappears, the whole region gets brightened. We speculate these brightenings are due to small-scale reconnections. Finally, the minfilament erupts as a blow out jet. The jet-based bright point is shown by yellow arrow in Figure 1(d), (h) and (l). Our finding of minifilament as a source of the jet is consistent with earlier findings [10,21,28,30,31].

\subsection{Underlying Photospheric Magnetic Field Evolution of Minifilament}
The minifilament formation in the filament channel is expected to be closely related to its magnetic field configuration. Due to lack of coronal and chromospheric magnetic field measurement, we do not get any direct information about magnetic field in the filament. However underlying photospheric magnetic field and its evolution are expected to provide us valuable information regarding the changes in magnetic field in the filament. In large-scale filament formation photospheric magnetic flux cancellation plays an important role [9,17,18]. Formation of small-scale filaments in the region of magnetic flux cancellation is also reported [9,13,34].

 In figure 2 we overplot the magnetic field contour from HMI on the AIA 171 image. This figure depicts that the minifilament is situated in neutral line of positive and negative magnetic field.
 
  In Figure 3 we represent three  images of HMI magnetogram subsequently one after the other at time 23:21:28.9 UT, 23:43:59.0 UT and 23:58:14.0 UT of $23^{rd}$ March 2017 respectively. We encircled our region of interest with an orange circle. In left panel the negative patch of magnetic field P is marked with a blue arrow and the negative patch of magnetic field N is marked by a red arrow. In the subsequent panels we see that those encircled negative and positive magnetic patches almost disappear since they cancel each other.
 
In figure 4 the rectangular black box shows our selection of the region for calculation of magnetic field evolution. We calculate the average positive magnetic field in this region and multiply with the area to obtain the positive magnetic flux in the selected region. The Figure 5 delineate the change in positive magnetic flux in our selected region with time (from 16:00 UT to 23:00 UT). Initially, we observe both increase and decrease of positive magnetic flux. But just prior to initiation of the eruption of minifilament, we find a continuous decrease in overall positive magnetic flux. The red vertical straight line shows the timing of minifilament eruption (approximately 22:08 UT). The decrease of positive magnetic flux continues as long as the eruption process of the minifilament goes on, and then it finally disappears. The increase in positive magnetic flux indicates new positive flux emergence. On the other hand decrease in positive magnetic flux could be attributed to cancellation of positive magnetic flux with negative magnetic flux. 

Flux linkage process is proposed as a mechanism of typical filament formation [16,20]. In flux linkage models two unconnected bipoles undergo reconnection due to new flux emergence or flux cancellation at photospheric magnetic field. Thus filament forms and evolves due to photospheric magnetic flux cancellation and new flux emergence. The convergence of opposite polarity magnetic patches is one of the basic aspect of various filament formation models in addition to flux linkage [5,6,15,32,33].

\section{Summary and conclusions}
We have studied an interesting case of a blowout jet from minifilament eruption on 23$^{rd}$ March of 2017. The minifilament forms at 15:50 UT in a quiet region towards the north-east of the solar disk. We have studied the evolution of minifilament and the evolution of magnetic field beneath the minifilament.  The minifilament disappear, and then it reappears, and finally, it erupts and forms a blowout jet. We detect the full and partial appearance and disappearance of the minifilament in AIA 171 ~\AA \hspace{0.5mm}, 193 ~\AA \hspace{0.5mm} and 304 ~\AA \hspace{0.5mm} wavelength channels. Finally, it gets destabilized, and its eruption initiates approximately at 22:08 UT. The eruption leads to the blow out jet. Our finding of minifilament eruption as a cause of jet is consistent with earlier findings [10,21,22,28,30,31].
 However in this work, we present a detailed account of the event which makes it unique.  \\\
We study the evolution of photospheric magnetic field beneath the minifilament. Initially, we observe a subsequent increase and decrease of positive magnetic flux beneath the minifilament region. Prior to initiation of the eruption of the minifilament, we observe a continuous decrease of positive magnetic flux which continues till its eruption and final disappearance. The increase and subsequent decrease in positive magnetic flux could be explained by new flux emergence and cancellation of magnetic flux respectively.  Our findings of flux cancellation in the region of minfilament formation are consistent with earlier findings [9,13,34].
\section{Acknowledgement}
R.M. acknowledge the University Grant Commission (UGC) and CESSI, IISER KOLKATA for funding.
 \bigskip\subsubsection*{REFERENCES}

 {\small
 \quad ~[1] Adams, M., Sterling, A. C., Moore, R. L., and Gary, G. A. 2014, Astrophys. J., 783, 11

[2] Alexander, D., and Fletcher, L. 1999, Solar Phys., 190, 167

[3] Canfield, R. C., Reardon, K. P., Leka, K. D., et al. 1996,
Astrophys. J., 464, 1016

[4] Chae, J., Qiu, J., Wang, H., and goode, P. R. 1999, As-
trophys. J. Lett., 513, L75

[5] DeVore, C. R., and Antiochos, S. K. 2000, Astrophys. J.,
539, 954

[6] Gaizauskas, V., Zirker, J. B., Sweetland, C., and Kovacs,
A. 1997, Astrophys. J., 479, 448

[7] Gilbert, H. R., Holzer, T. E., Burkepile, J. T., and Hund-
hausen, A. J. 2000, Astrophys. J., 537, 503

[8] Gopalswamy, N., Shimojo, M., Lu, W., et al. 2003, As-
trophys. J., 586, 562

[9] Hermans, L. M., and Martin, S. F. 1986, in NASA Confer-
ence Publication, Vol. 2442, NASA Conference Publica-
tion, ed. A. I. Poland

[10] Hong, J., Jiang, Y., Zheng, R., et al. 2011, Astrophys.
J. Lett., 738, L20

[11] Huang, Z., Madjarska, M. S., Doyle, J. G., and Lamb, D. A.
2012, Astron. Astrophys., 548, A62

[12] Jing, J., Yurchyshyn, V. B., Yang, G., Xu, Y., and Wang,
H. 2004, Astrophys. J., 614, 1054

[13] Lee, S., Yun, H. S., Chae, J., and Goode, P. R. 2003, Jour-
nal of Korean Astronomical Society, 36, S21

[14] Lemen, J. R., Title, A. M., Akin, D. J., et al. 2012, Solar
Phys., 275, 17

[15] Litvinenko, Y. E., and Martin, S. F. 1999, Solar Phys., 190,
45

[16] Martens, P. C., and Zwaan, C. 2001, Astrophys. J., 558,
872

[17] Martin, S. F. 1998, Solar Phys., 182, 107

[18] Martin, S. F., Livi, S. H. B., and Wang, J. 1985, Australian
Journal of Physics, 38, 929

[19] Moore, R. L., Tang, F., Bohlin, J. D., and Golub, L. 1977,
Astrophys. J., 218, 286

[20] Panesar, N. K., Innes, D. E., Schmit, D. J., and Tiwari,
S. K. 2014, Solar Phys., 289, 2971

[21] Panesar, N. K., Sterling, A. C., and Moore, R. L. 2017,
Astrophys. J., 844, 131

[22] Panesar, N. K., Sterling, A. C., Moore, R. L., and Chakra-
pani, P. 2016, Astrophys. J. Lett., 832, L7

[23] Raouafi, N. E., Patsourakos, S., Pariat, E., et al. 2016,
Space Sci. Rev., 201, 1

[24] Ren, D. B., Jiang, Y. C., Yang, J. Y., et al. 2008, As-
trophys. Space Sci., 318, 141

[25] Sakajiri, T., Brooks, D. H., Yamamoto, T., et al. 2004,
Astrophys. J., 616, 578

[26] Scherrer, P. H., Schou, J., Bush, R. I., et al. 2012, Solar
Phys., 275, 207

[27] Schou, J., Scherrer, P. H., Bush, R. I., et al. 2012, Solar
Phys., 275, 229

[28] Shen, Y., Liu, Y., Su, J., and Deng, Y. 2012, Astrophys.
J., 745, 164

[29] Shibata, K., Ishido, Y., Acton, L. W., et al. 1992, Pub.
Astron. Soc. Japan, 44, L173

[30] Sterling, A. C., Moore, R. L., Falconer, D. A., and Adams,
M. 2015, Nature (London), 523, 437

[31] Sterling, A. C., Moore, R. L., Falconer, D. A., et al. 2016,
Astrophys. J., 821, 100

[32] van Ballegooijen, A. A., and Cranmer, S. R. 2010, Astro-
phys. J., 711, 164

[33] van Ballegooijen, A. A., and Martens, P. C. H. 1989, As-
trophys. J., 343, 971

[34] Wang, J., Li, W., Denker, C., et al. 2000, Astrophys. J.,
530, 1071

[35] Wang, Y.-M., Sheeley, Jr., N. R., Socker, D. G., et al.
1998, Astrophys. J., 508, 899

[36] Young, P. R., and Muglach, K. 2014, Solar Phys., 289, 3313

[37] Zuccarello, F., Battiato, V., Contarino, L., Romano, P.,
and Spadaro, D. 2007, Astron. Astrophys., 468, 299

 }

\end{document}